\documentclass{sigchi-ext}
\usepackage[T1]{fontenc}
\usepackage{textcomp}
\usepackage[scaled=.92]{helvet} 
\usepackage{graphicx} 
\usepackage{balance}  
\usepackage{booktabs} 
\usepackage{ccicons}  
\usepackage{ragged2e} 

\def\plaintitle{Advancing Inclusive Design in Multiple Dimensions} \def\plainauthor{Gabriela Molina Leon}

\def\plainkeywords{knowledge production; participatory design; multimodal interaction; data visualization}

\title{Advancing Inclusive Design in Multiple Dimensions}

\numberofauthors{1}
\author{%
  \alignauthor{%
    \textbf{Gabriela Molina León}\\
    \affaddr{University of Bremen} \\
    \affaddr{Germany} \\
    \email{molina@uni-bremen.de} } }

\definecolor{linkColor}{RGB}{6,125,233}
\hypersetup{%
  pdftitle={\plaintitle},
  pdfauthor={\plainauthor},
  pdfkeywords={\plainkeywords},
  bookmarksnumbered,
  pdfstartview={FitH},
  colorlinks,
  citecolor=black,
  filecolor=black,
  linkcolor=black,
  urlcolor=linkColor,
  breaklinks=true,
}

\begin{document}

\CopyrightYear{2023}
\setcopyright{rightsretained}
\conferenceinfo{CHI'23,}{April  23--28, 2023, Hamburg, Germany}
\isbn{xxx}
\doi{https://doi.org/10.1145/3334480.XXXXXXX}
\copyrightinfo{\acmcopyright}

\maketitle

\RaggedRight{} 

\begin{abstract}
   Making technology design inclusive requires facing multiple challenges in different dimensions: the populations we work with, who we are, what interaction possibilities we consider, and what context we examine. We reflect on these challenges and propose two main measures to achieve research inclusiveness.
\end{abstract}




\section{Introduction}
We draw from recent research on knowledge production, participatory design, and data visualization to reflect on the challenges researchers face to make technology design inclusive. We discuss four dimensions of the problem and propose two ways to combat bias and explore the limits of interaction design research:
\begin{enumerate}\compresslist%
    \item Think of you and your sample. 
    \item Consider diverse interaction modalities and devices.
\end{enumerate}

\section{Think of you and your sample}
The recent study of Linxen et al.~\cite{linxen21} made clear that human-computer interaction research (HCI) has mainly focused on people who are Western, Educated, Industrialized, Rich, and Democratic (WEIRD). The implication is that the field makes claims about human behavior and preferences based on experiments conducted with participants representing less than 12\% of the world population.
To overcome the WEIRD bias, we must reach out to the populations in the margins. These margins can be socio-economical or based on individual characteristics such as age, race, and gender. We need to consider all aspects and always ask ourselves if these populations have been taken into account in our work. Designing for diverse user groups benefits from a participatory mindset. Participatory design~\cite{lee18} has often been used for designing with specific audiences, such as senior citizens~\cite{bull17}. Participatory methods such as co-creation~\cite{sanders08} help to involve and empower participants, enabling the potential users to shape the design of their tools.

However, bias does not apply only to research participants but also to the researchers themselves. In the top conferences and journals, authors usually come from WEIRD institutions. As the Global North provides more infrastructure and resources to conduct research projects, the people shaping the research agenda are usually located there. Accordingly, the venues are in the Global North, and consequently, this enforces the systematic exclusion of researchers who cannot afford to attend, given financial and travel limitations.
To overcome these challenges, we should bring the research to the Global South. As the infrastructure to conduct hybrid events has further developed during the COVID-19 pandemic, we should take advantage of it to reach out to researchers from other world regions. Recent efforts on the topic of citational justice~\cite{kumar21} have proved to be successful by organizing workshops in regional conferences, such as CLIHC (the Latin American Conference on HCI)~\cite{cjc21} and India HCI, to learn more from the researchers and communities of Latin America and South Asia.

\subsection{Social norms matter}
Within the Global South and the Global North, we can also find specific country-based patterns. In a recent qualitative study, Naveed et al.~\cite{naveed22} found that religion and gender strongly influence the privacy concerns of mobile device owners in Pakistan. Privacy is a critical factor to consider for the design of conversational user interfaces (CUIs). In a study conducted in Germany, Seiderer et al.~\cite{Seiderer20} investigated the requirements of senior citizens for using voice assistants. They prioritized \emph{privacy-by-design}, which fits the importance German citizens give to privacy in the context of technology. Thus, we should present such findings making clear the context in which the research took place.

\section{Consider diverse modalities and devices}
Speech interaction has been increasingly incorporated in data visualization research projects (e.g.~\cite{kim21, srinivasan20}), often as part of multimodal systems. Multimodal interaction has been a subject of study for a few decades in the field of HCI~\cite{oviatt99}.
Given that working with visualizations classically relies on vision, multimodal interaction has the potential to make visualizations more accessible. Researchers are exploring new ways to support interactivity for people who have visual impairments. That is a critical milestone to make visualization design inclusive, given that at least two billion people worldwide have a visual impairment~\cite{who19}.

Speech recognition for multimodal interaction has also been tested in different devices and settings. Kim et al.~\cite{kim21} combined speech and touch to support mobile phone users in exploring their personal health data. Saktheeswaran et al.~\cite{saktheeswaran20} compared touch-only interaction, speech-only interaction, and touch-and-speech interaction for exploring network data on a large vertical display and found that people preferred multimodal interaction because the modalities complemented each other. When speech commands led to mistakes, participants corrected them through touch.
In a recent study on interacting multimodally with visualizations on tablets~\cite{molina22}, participants struggled to make specific speech queries because they were no English native speakers and were unsure about the pronunciation of country names they were not familiar with. Other interaction modalities can help them overcome these obstacles, e.g.,~using touch to select the country name based on a list of suggestions generated via speech. 
However, participants actually preferred to use pen interaction over touch and speech on the tablet.
Thus, exploring interaction modalities across different devices may lead to different results regarding user behavior and preferences.

\section{Conclusion}
We have explored only four dimensions to consider for the inclusive design of CUIs. While there are more aspects to consider, we hope these reflections help continue the discussion and search for solutions to make research more inclusive.

\section{Acknowledgements}
This work was funded by the Deutsche Forschungsgemeinschaft (DFG, German Research Foundation)—project number 374666841—SFB 1342.

\balance{} 

\bibliographystyle{SIGCHI-Reference-Format}
\bibliography{sample}

\end{document}